\documentclass[prl,twocolumn,showpacs]{revtex4}

\usepackage{times}
\usepackage{bm}
\usepackage[dvips]{graphicx}
\usepackage{amsmath}
\usepackage{bm}
\usepackage{natbib}
\usepackage{color}
\usepackage{epstopdf}
\begin{document}
\title{Anomalous Ballistic Transport in Disordered Bilayer Graphene:\\
 Dimer Vacancies induced Dirac Semimetal }
 \author{Dinh Van Tuan,$^{1}$   and Stephan  Roche$^{1,2}$}
 \affiliation{$^1$ICN2 - Institut Catala de Nanociencia i Nanotecnologia,
 Campus UAB, 08193 Bellaterra (Barcelona), Spain\\
 $^2$ICREA, Instituci\'{o} Catalana de Recerca i Estudis Avan\c{c}ats,
  08070 Barcelona, Spain}
\date{\today}
\begin{abstract}
We report anomalous quantum transport features in bilayer graphene in presence of a random distribution of structural vacancies. By using an efficient real-space Kubo-Greenwood transport methodology, the impact of a varying density of dimer versus non-dimer vacancies is investigated in very large scale disordered models. While non-dimer vacancies are shown to induce localization regimes, dimer vacancies result in an unexpected ballistic regime whose energy window surprisingly enlarges with increasing impurity density. Such counterintuitive phenomenon is explained by the formation of an effective linear dispersion in the bilayer bandstructure, which roots in the symmetry breaking effects driven by dimer vacancies, and provides a novel realization of Dirac semimetals in high dimension. 
\end{abstract} 

\maketitle

 \textit{Introduction}.- 
Single layer graphene (SLG) has attracted a great attention owing to its remarkable electrical, chemical and mechanical properties providing an endless list of novel opportunities for practical applications \cite{Neto2009}.  SLG possesses a linear electronic spectrum with chiral (A-B sublattice) symmetry which lead to exotic low-energy transport such as Klein tunneling \cite{Katsnelson2006, Young2009}, weak antilocalization \cite{McCann2006, Tikhonenko2009}, half-integer quantum Hall effect \cite{Novoselov2005, Zhang2005}.   

Bilayer graphene (BLG) differs from SLG by the parabolic band dispersion which however retains the chiral nature of low-energy electronic excitations. One of the salient and unique property of BLG is the possibility of creating an electronic bandgap by applying external gate voltage \cite{McCann20061, Castro2007}. However surprisingly, the understanding of quantum transport in disordered BLG remains far less understood than the SLG case, because of the enhanced structural complexity. Experimental studies evidence critical differences in transport behaviors of SLG and BLG \cite{Bouchiat}. Some generalization of the localization theory in BLG has been derived \cite{BLGWL}, while a minimum conductivity $\sigma_{min} \simeq 4e^2/h$ is predicted at the charge neutrality point (CNP) \cite{Katsnelson2007,Sarma2010,Yuan2010}. 
A recent scanning tunneling microscopy (STM) study shows that vacancies in graphite induce peculiar impurity states known as zero-energy modes (ZEMs) which are maximally localized at the defect position and then decay as the inverse of the distance from the vacancy \cite{Ugeda2010}. The impact of ZEMs on BLG is so far poorly understood, especially the role played by dimer and non-dimer vacancies which strongly differ in terms of symmetry breaking characteristics.

In this Letter, we start by analysing the localization features of ZEMs in BLG for all types of vacancies, and found a highly inhomogenous sublattice state population (pseudospin polarization) as reported in STM experiments \cite{Ugeda2010}. The depletion of low energy states in one sublattice is additionally further classified into two different classes depending on the vacancy position. Then by using efficient computational methods, we explore quantum transport in disordered BLG and analyze how the local nature of the vacancy (dimer versus non-dimer) impact on scattering and localization phenomena. The type and concentration of vacancies are found to dictate the nature of the transport regime which ranges from weak localization to anomalous ballistic conduction.
 
\begin{figure}[htbp]
 \includegraphics[width=0.5\textwidth]{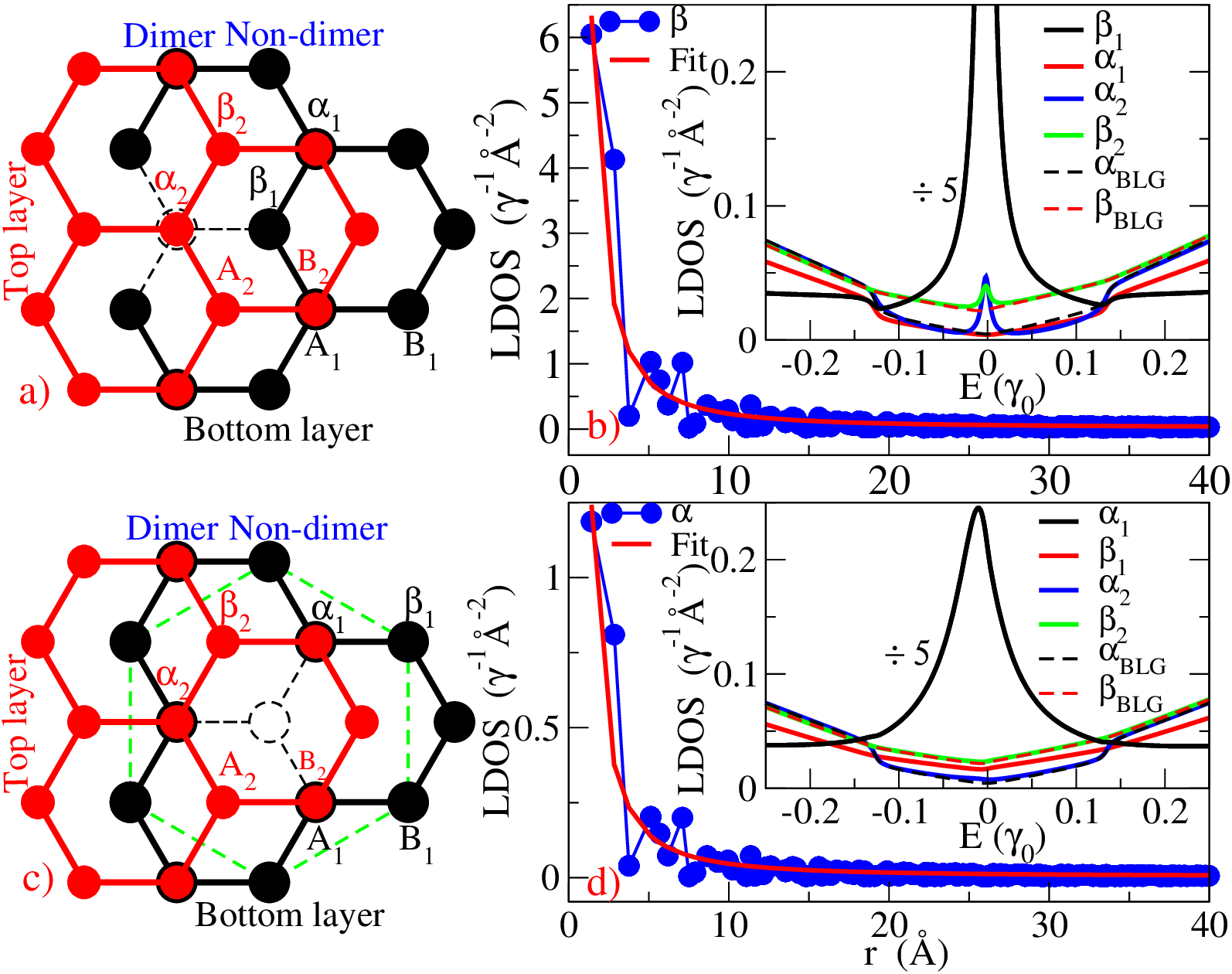}
\caption{(color online): Sketch of a dimer (a) and non-dimer (c) vacancy (open dashed circle) in a BLG composed of a top (in red) and a bottom (in black) layers. State distribution around a dimer (b) and a non-dimer (d) vacancy on its layer in the opposite sublattice.  Power law fit (red solid curve). Inset: LDOSs shown in (a) and (c), closest to dimer (b) and non-dimer vacancies (d), together with the pristine BLG (dashed curves).}
 \label{Fig1}
 \end{figure}

\textit{Electronic features induced by vacancies in graphene}.-BLG (Fig.\ref{Fig1}(a) and (c)) can be considered as two coupled  SLGs with the top layer (in red)  shifted a carbon bond  from the bottom layer (in black). Consequently, BLG consists of four carbon atoms in its unit cell, two carbons $A_1,B_1$ in the bottom SLG unit cell and $A_2, B_2$ in the top layer  where $B_2$ lies on the top of $A_1$, namely dimer or $\alpha$ sites whereas $B_1, A_2$ are called non-dimer or $\beta$ sites.  The tight-binding Hamiltonian model for BLG reads \cite{McCann2013,Jung2014}
 \begin{eqnarray}
&{\mathcal{H}}&=-\gamma_0\sum_{\langle ij\rangle,l=1,2 }a_{l,i}^+b_{l,j}+\gamma_1\sum_{ i }a_{1,i}^+b_{2,i}
+\gamma_3\sum_{ i,i' }b_{1,i}^+a_{2,i'}\\
&-&\gamma_4\sum_{ i,j' }\left\{a_{1,i}^+a_{2,j'}+b_{1,i}^+b_{2,j'}\right\}+h.c. 
+\Delta\sum_{ i }\left\{a_{1,i}^+a_{1,i}+b_{2,i}^+b_{2,i}\right\} \nonumber
\label{HamilBLG}
\end{eqnarray}
where $l=1,2$ labels the bottom and top layer respectively. The annihilation (creation) operators acting on $A_1$, $B_1$, $A_2$, $B_2$ are denoted
$a_1$, $b_1$, $a_2$, $b_2$ ($a_1^+$, $b_1^+$, $a_2^+$, $b_2^+$).  The first term in $\mathcal{H}$ describes the intralayer hopping  between nearest-neighbor $\pi$-orbitals. The second term denotes the interlayer hopping ($\gamma_1=340$ meV) between dimer sites $\{A_1,B_2\}$, while the third term gives the interlayer coupling between $B_1$ and its closet $A_2$ with $\gamma_3=280$ meV. The fourth term corresponds to the hopping from $A_1$ to its nearest $A_2$ site and from $B_1$ to its nearest $B_2$ neighbors ($\gamma_4=145$ meV). The energy asymmetry between dimer and non-dimer sites is taken into account by introducing in the final term of $\mathcal{H}$ an on-site energy difference $\Delta=9.6$ meV between dimer sites. All these parameters are derived from the ab-initio calculations \cite{Konschuh2012,Jung2014}.

The Local Density of States (LDOS) of pristine BLG are first scrutinized (dash curves in the inset of Fig.\ref{Fig1}(b) or (d)). Since the two layers are identical one can restrict the discussion to $\alpha$ (dimer) and $\beta$ (non-dimer) sites. The LDOSs on both sites show a sudden change in the slope at $E=\pm\gamma_1$, especially for $\alpha$ sites (black dashed line) owing to the contribution of higher energy bands \cite{McCann2013}. The LDOSs also clearly exhibit fingerprints of {\it pseudospin polarization} on each layer, in the sense that the state mainly populate $\beta$ sites (${\rm LDOS}_{\beta}\neq 0$ and ${\rm LDOS}_{\alpha}= 0$).

A monovacancy on BLG is simulated by removing a carbon atom from the pristine BLG which can be either chosen on the dimer or the non-dimer site. Fig.\ref{Fig1}(a) and (c) show the sketch of BLG with dimer and non-dimer vacancies (dashed circle), while the decay of the impurity states are given in Fig. \ref{Fig1}(b) and (d), respectively. We first consider the effect of a single vacancy on its layer (here the bottom layer). Our results (Fig.\ref{Fig1}(b) and (d) inset) show that both dimer and non-dimer vacancies have a strong impact on the layer where the vacancy resides with a strong depletion of ZEM on the vacancy sublattice and a state abundance in the other sublattice (similarly to SLG, see Supp. Mat.\cite{Dinh2015Sub}). This is seen in Fig.\ref{Fig1} (b) and (d) (insets)  by comparing the LDOSs of the two nearest sites in the bottom layer (black and red solid lines) with the pristine BLG case (dashed lines).  Dimer vacancies also more strongly impact at high energy as seen in the decay of the LDOS at $E=\pm \gamma_1$. The spatial distribution of the sublattice abundant state induced by the vacancy in its layer is further investigated in Fig.\ref{Fig1}(b) and (d) where the data (closed circle) for both dimer (Fig.\ref{Fig1} (b)) and non-dimer vacancy (Fig.\ref{Fig1} (d)) are fitted by power law $r^{-2}$  (red lines). The depletion of the charge density in the vacancy sublattice can be actually classified into two different classes.  The first class involves six second-nearest neighbors of the vacancy and forms a hexagonal lattice (green dashed line in Fig.\ref{Fig1}(c)) with an enlarged graphene lattice constant by a factor of $\sqrt{3}$, whereas the second class together with the vacancy populate the centers of the previous hexagons.  Such long range nature of the impurity state distribution will be key in understanding anomalous ballistic transport. We next investigate the effect of a single BLG vacancy on its adjacent layer.  While the dimer vacancy introduces ZEMs on both sublattices of the adjacent layer (the peaks at $E = 0$ in green and blue curves in Fig.\ref{Fig1}(b) inset), the non-dimer vacancy leaves the second layer almost unaffected. Indeed, the LDOSs for $\alpha_2$ (blue solid line) and $\beta_2$ (green solid line) site in Fig.\ref{Fig1}(d) inset are almost identical to the ones for pristine BLG (dashed lines). 

\textit{Transport properties in the dilute vacancy limit}.-Charge transport properties of BLG are investigated for a finite density of vacancies, differentiating the cases with only dimer or non-dimer vacancies (uncompensated cases) from the case with equally distributed mixture of both types (compensated case). Here we also assume an equal distribution among top and bottom layers. We use a real-space order-N wave packet evolution approach \cite{RocheKG}. The Kubo-Greenwood conductivity is written as $\sigma (E,t)= e^{2} \rho(E) \Delta X^{2}(E,t)/ t$, where $\rho(E)$ is the DOS and $\Delta X^{2}(E,t)$ is the mean quadratic displacement of the wave packet and gives the diffusion coefficient $D(E,t)=\Delta X^{2}(E,t)/t$. With disorder, $D(t)$ changes from a ballistic motion to a saturation regime, from which the mean free path $\ell_e$ is deduced through $\ell_e(E)=D^{\text{max}}(E)/2 v(E)$ ($v(E)$ is the velocity and $D^{\text{max}}$ the maximum value). At long times, $D(E,t)$ eventually decay owing to quantum interferences which drive the system either to weak or strong (Anderson) localization regime \cite{Cresti2013}.

\begin{figure}[htbp]
\includegraphics[width=0.5\textwidth]{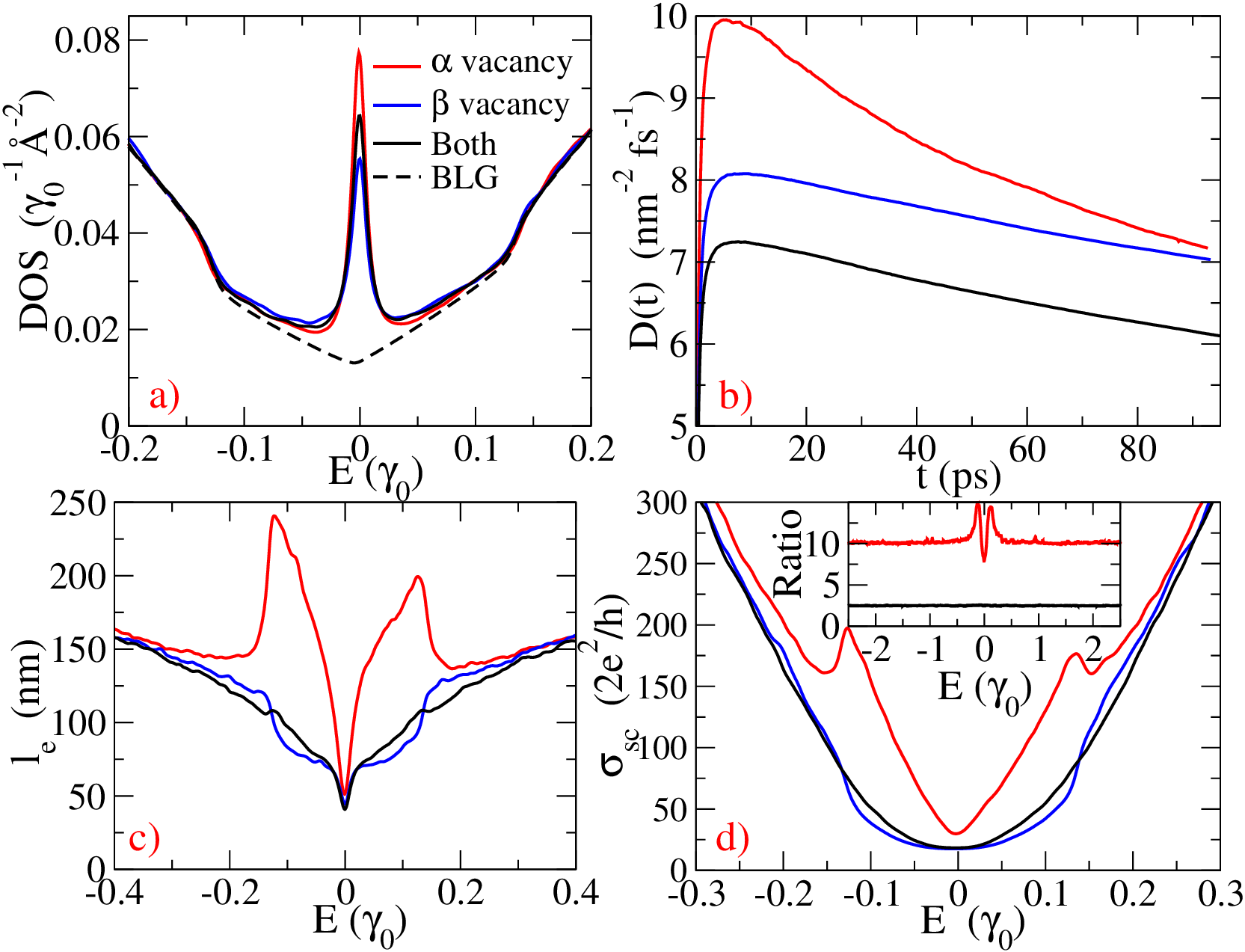}
\caption{(color online)(a) Total DOS for BLG samples with 0.05\% vacancies together with the pristine DOS (dashed curve), with only dimer vacancies (red curve), non-dimer vacancies (blue curve) or a mixed case (black curve). (b), (c) and (d) $D(E=0,t)$, $\l_e(E)$ and $\sigma_{sc}(E)$ corresponding to the DOSs in (a). (d, inset) $\sigma_{sc}(0.02\%)/\sigma_{sc}(0.05\%)$ (black curve) and $\sigma_{sc}(0.05\%)/\sigma_{sc}(0.5\%)$ (red curve).}    
\label{Fig2}
\end{figure}

Fig.\ref{Fig2}(a) shows the total DOS for BLG with 0.05\% of dimer vacancies $\alpha$ (red solid curve), non-dimer $\beta$ (blue solid curve) and mixed $\alpha-\beta$ case (black solid curve), together with the total DOS for pristine case (dashed curve). As already suggested in Fig.\ref{Fig1} (b) and (d) (insets), especially at the CNP, dimer vacancies more strongly affect the DOS than non-dimer ones, whereas the depletion of state at $E=\pm \gamma_1$ is unobservable. However, a marked difference is observed on the mean free path $\l_e$ (Fig.\ref{Fig2}(c)) and semiclassical conductivities
$\sigma_{sc}=e^{2} \rho(E)D^{\text{max}}$ (Fig.\ref{Fig2}(d)) with two significant peaks at $E=\pm \gamma_1$. The values of $\l_e$ (resp. $\sigma_{sc}$) at such energy are higher for the dimer than for the non-dimer vacancies by a factor of 3 (resp.  a factor of 2) indicating a stronger scattering efficiency for non-dimer vacancies. This likely originates from the imbalance of state at $E=\pm \gamma_1$ on the two sublattices around the non-dimer vacancies (Fig.\ref{Fig1}(b),(d) insets). The stronger impurity scattering strength at high energy for non-dimer vacancies is also confirmed by the fact that $\l_e$ and $\sigma_{sc}$ are almost the same for the compensated case (black solid curves in Fig.\ref{Fig2}(c),(d)) and the non-dimer vacancy case (blue solid curves).  For three cases  $\l_e$ are almost the same but $\sigma_{sc}$ for the dimer vacancy is larger due to enhanced DOS. $D(E,t)$ further evidence much stronger localization effects for the dimer vacancies  (Fig.\ref{Fig2}(b)), consistently with the DOS results (Fig.\ref{Fig1} (b) and (d)). Importantly in the low impurity density limit ($\leq 0.05\%$), regardless the nature of vacancies, a localization regime is obtained for the whole energy spectrum. 

However, surprising features are obtained for high enough vacancy density when the affected spatial areas around vacancies start to overlap. One first scrutinizes the ratios of semiclassical conductivity $\sigma_{sc}$ for the compensated case in the dilute limit with 0.02\%, 0.05\% of vacancy, and dense limit with  0.5\% of vacancies (Fig.\ref{Fig2}(d) (inset)). In the dilute limit $\sigma_{sc}$ perfectly obeys the Fermi's golden rule $\sim 1/n_i$ in the whole energy band (black solid curve in Fig.\ref{Fig2}(d) (inset).  The Fermi's golden rule, however, underestimates (overestimates) the values of $\sigma_{sc}$ at CNP ($E=\pm \gamma_1$) in the dense impurity limit (red solid curve in Fig.\ref{Fig2}(d) (inset), i.e. when increasing the vacancy coverage $\sigma_{sc}$ decreases faster at $E=\pm \gamma_1$ and slower at the CNP. 

\begin{figure}[htbp]
\includegraphics[width=0.5\textwidth]{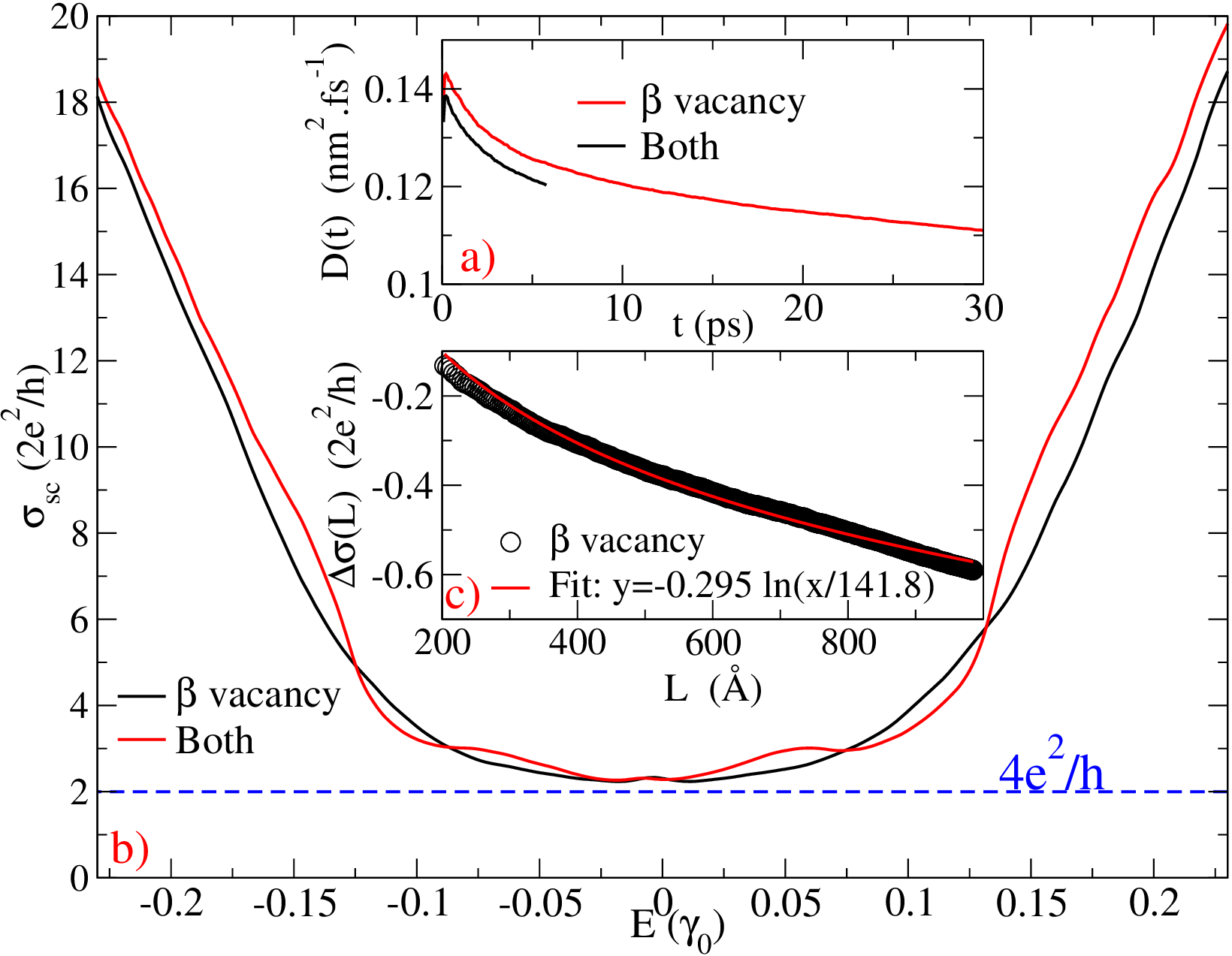}
\caption{(color online) $D(t)$ (a) and $\sigma_{sc}$ (b) for 0.5\% vacancy in the non-dimer (red curves) or compensated case (black curves),  with the minimum conductivity $4e^2/h$ for BLG (dashed curve). (c) $\Delta\sigma(L)=\sigma(L)-\sigma_{sc}$ for $ 0.5\%$ of non-dimer vacancy (open circle) and the fit to weak localization theory (red solid curve). }
\label{Fig3}
\end{figure}

\begin{figure}[htbp]
\begin{center}
\leavevmode
\includegraphics[width=0.5\textwidth]{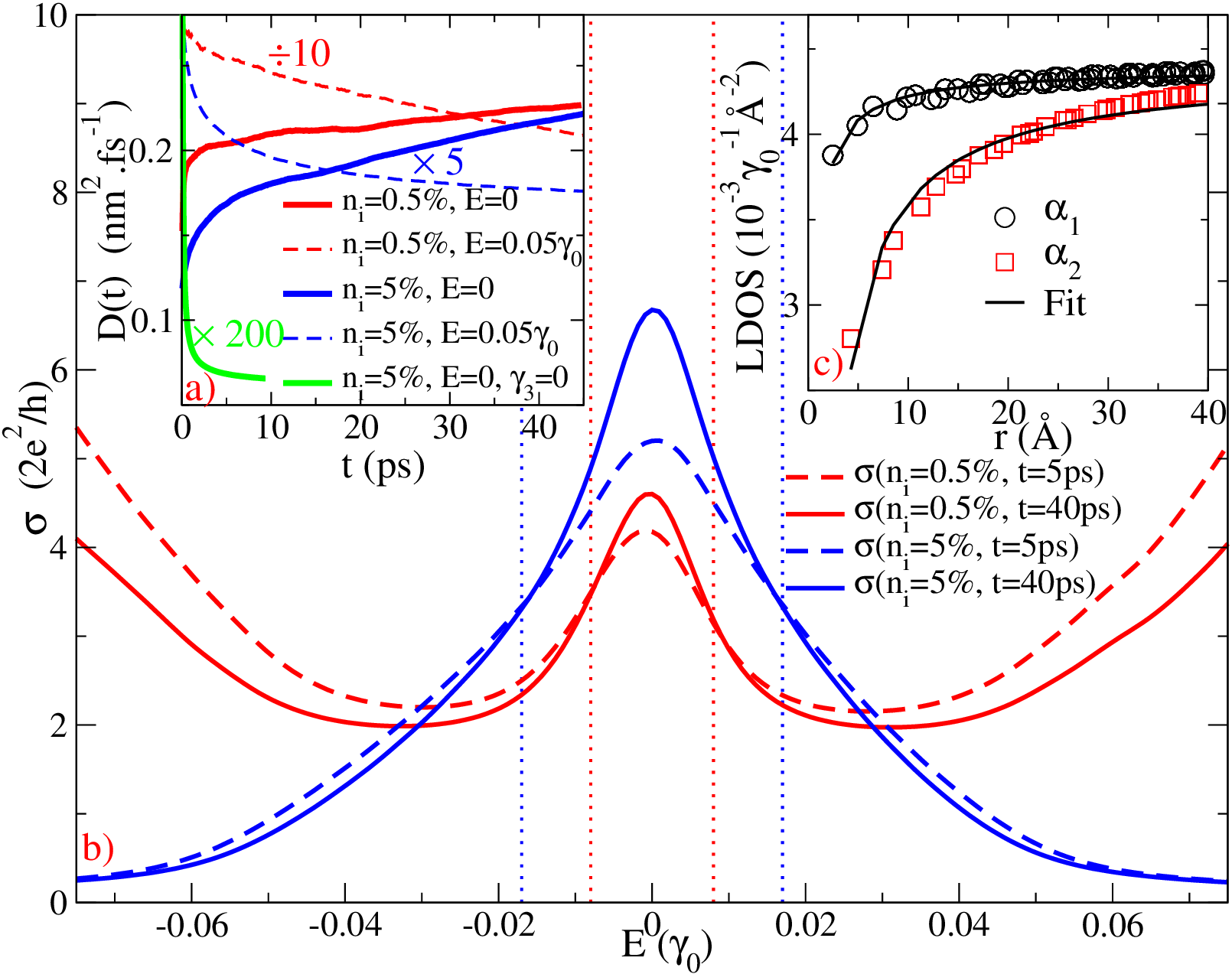}
\caption{(color online) (a) $D(E=0,t)$ (solid lines) and $D(E=0.05\gamma_0,t)$ (dashed lines) for 0.5\% (red lines) and 5\% (blue lines) of dimer vacancies. The green line gives the case of 5\% of dimer vacancy with $\gamma_3=0$.  (b) Conductivities at $t=5$ps (dashed lines) and at $t=40$ps (solid lines) for 0.5\% (red lines) and 5\% (blue lines) of dimer vacancies. (c) Charge distribution on the vacancy sublattice around a dimer vacancy, which separate into two classes (see text), together with their fittings to the power law (solid lines). }
\label{Fig4}
\end{center}
\end{figure}

\textit{Transport properties in dense vacancy limit}.-Fig.\ref{Fig3} and Fig.\ref{Fig4} show results for 0.5\% of non-dimer vacancies and compensated case (Fig.\ref{Fig3} ) and dimer vacancies (Fig.\ref{Fig4}). Charge transport in presence of non-dimer vacancies and for the compensated case similarly exhibit a localization behavior (as seen in Fig.\ref{Fig3}(a)). The low-energy semiclassical conductivities also clearly saturates at $ \sigma_{sc}^{min}\simeq 4e^2 /h$ (dashed curve) \cite{Yuan2010}.

The calculation of the quantum correction of the semiclassical conductivity $\Delta\sigma(L)=\sigma(L)-\sigma_{sc}$ at CNP (Fig.\ref{Fig3}(c)) further confirms that the non-dimer vacancies induce localization, since a very good fit $\Delta\sigma(L)\simeq a\frac{2e^2}{h} \ln(L/\l_e^*)$ is obtained with $a=-0.295\sim -1/\pi$ in agreement with 2D weak localization theory ($\Delta\sigma(L)\simeq- \frac{2e^2}{\pi h} \ln(L/\l_e^*)$). 

The case of high density of dimer vacancy is remarkably different from the non-dimer one. Fig.\ref{Fig4}(a) shows $D(E,t)$ for 0.5\% dimer vacancies at CNP (red solid line) and at high energy $E=0.05\gamma_0$ (red dashed line). While high energy charge transport quickly enters the localization regime (see $D(E=0.05\gamma_0,t)$), in sharp contrast $D(E=0,t)$ increases linearly without any sign of saturation, as expected in a truly ballistic motion.  This remarkable behavior is captured in the quantum conductivity scaling $\sigma(t)$ in the vicinity of CNP. Fig.\ref{Fig4}(b) shows that $\sigma(t)$ at $t=5$ ps (dashed lines) is smaller than its value at $t=40$ ps (solid lines) around CNP. A crossover from the ballistic regime to the localization regime actually occurs at a critical energy $E_c=0.008\gamma_0$ (red dotted lines). This anomalous behavior is also observed for a higher dimer vacancy density of $5\%$ (Fig.\ref{Fig4}(a, b) (blue  lines))., for which  a significant enlargement of the ballistic-transport energy range is obtained ($E_c=0.017\gamma_0$ (blue dotted lines)). This is a remarkably counterintuitive phenomenon in which the quantum conductivity increases with disorder density.  As explained below, the interlayer hopping between non-dimer sites $\gamma_3$ governs such anomalous behavior. First, by setting $\gamma_3=0$ for the case of $5\%$ dimer vacancies, $D(t)$ is seen to exhibit a strong localization with a very fast decay at long times and a value reduced by two orders of magnitude when compared to the case with $\gamma_3 \neq 0$ (Fig.\ref{Fig4}(a) (green solid line)). 
 
To rationalize such effect, let us then consider a simple tight-binding model by suppressing the terms $\gamma_4$ and $\Delta$ in Eq.(\ref{HamilBLG}), which is equivalent to the continuum model \cite{McCann20061,McCann2013}
\begin{equation}
\hat{H}=-\frac{1}{2m}
\left( \begin{array}{cc}
0 & (\pi^\dagger)^2 \\
 (\pi)^2 & 0 
  \end{array} \right)
  +\xi v_3
  \left( \begin{array}{cc}
0 & \pi \\
 \pi^\dagger & 0 
  \end{array} \right)
  \label{Continuum}
\end{equation} 
It describes two possible {\it interlayer hopping} events between non-dimer sites, the ones with dominant charge density (Fig.\ref{Fig1}(b) and (d) inset (dashed lines)). The first term activates hopping via dimer sites which generate a mass term $m=\gamma_1/2v^2 $ and a parabolic band \cite{McCann20061}. The second term describes a direct hopping $\gamma_3$ between non-dimer sites in the bottom and top layers with velocity $v_3=(\sqrt{3}/2)a\gamma_3/\hbar$. This term is dominant at low energy and generates a linear energy dispersion \cite{McCann20061,McCann2013}
\begin{equation}
\epsilon=\sqrt{(v_3 p)^2-\frac{\xi v_3p^3}{m}   \cos(3\phi)+\left(\frac{p^2}{2m}\right)^2}
\end{equation}
where $\xi=+1$ ($-1$) corresponds to the  $K$  ($K'$) valley and $\phi $ labels the momentum direction. This linear dispersion occurs for $\epsilon<\frac{1}{2}\gamma_1 (v_3/v)^2$.  Hereafter we will show that this value is enhanced in presence of dimer vacancies and yield anomalous ballistic motion.

The impurity state created by a dimer vacancy is mainly localized on non-dimer sites, and has almost no weight on dimer sites around the defect. Such sublattice polarization induced around the vacancy is shown in Fig.\ref{Fig1}(b). Fig.\ref{Fig4} (c) further evidences the depletion of LDOS on the dimer sites of the layer where the vacancy is lying.  Similarly to the SLG case (see \cite{Dinh2015Sub}), the LDOSs on dimer sites can also be separated into two classes (circle and square symbols) which are both well fitted with the scaling law $\rho=\rho_0-b r^{-0.9}$ (black solid lines) where $\rho_0$ is the LDOS far away from the vacancy.  From Eq.(\ref{Continuum}), one observes that the abundance of charge density on the non-dimer sites leads to the enhancement of the second term in Eq.(\ref{Continuum}) which corresponds to the direct hopping between non-dimer sites.  The reduction of electronic state on the dimer sites, on the other hand, leads to a reduction of the first term in Eq.(\ref{Continuum}). Indeed, this term involves a hopping process in which electron has to hop into a dimer site. This process is limited by dimer vacancies because it creates around it, in dimer sites, a wide area of depletion of electron density, i.e. prohibiting electron residence on the dimer sites in the vacancy vicinity.   We can thus consider BLG with a high enough density of dimer vacancies as an effective pristine BLG with renormalized parameters $v^*<v$ and $v_3^*>v_3$. The more dimer vacancies in the BLG, the smaller the renormalized value of $v^*$ and the larger the renormalized value of $v_3^*$. This thus leads to the expansion of the energy range ($\frac{1}{2}\gamma_1 (v_3^*/v^*)^2$ )  where the linear dispersion dominates.  This also explains the increase of critical energy $E_c$ with the dimer vacancy density in Fig.\ref{Fig4}(b). Such chiral electrons in the renormalized linear BLG energy dispersion share many similarities with low-energy excitations propagating in SLG (the sublattice index is replaced by the layer index) yielding backscattering suppression. This situation thus provides a novel realization of a Dirac semimetal in high dimensionality, which is an unconventional transport regime provoked by symmetry effects of impurities.  This could motivate a systematic exploration of quantum transport in irradiated BLG, following pioneering experimental studies \cite{Ugeda2010}.

\end{document}